\documentclass[a4paper,10pt]{article}
\usepackage[utf8x]{inputenc}
\usepackage{graphicx}
\usepackage{amsmath}
\usepackage{cite}
\usepackage{authblk}
\usepackage{color}
\usepackage[margin=3cm]{geometry}

\title{\bf Climate network and complexity approach predict neutral ENSO event for 2025}
\date{}
\renewcommand\Affilfont{\itshape\small}

\author[1] {Josef Ludescher}
\author[2] {Jun Meng}
\author[3] {Jingfang Fan}
\author[4] {Armin Bunde}
\author[5] {Hans Joachim Schellnhuber}

\tiny
\affil[1] {Potsdam Institute for Climate Impact Research (PIK), Member of the Leibniz Association, 14412 Potsdam, Germany}
\affil[2] {School of Science, Beijing University of Posts and Telecommunications, Beijing 100876, China}
\affil[3] {School of Systems Science, Beijing Normal University, 1000875 Beijing, China}
\affil[4] {Institute for Theoretical Physics, Justus Liebig Universit\"at Gießen, 35392 Gießen, Germany}
\affil[5] {International Institute for Applied Systems Analysis, Laxenburg 2361, Austria}

\begin{document}

\maketitle

\begin{abstract}
The El~Ni\~no Southern Oscillation (ENSO) is the strongest driver of interannual global climate variability and can lead to extreme weather events like droughts and flooding.
Additionally, ENSO influences the mean global temperature with strong El~Ni\~no events often leading, in a warming climate, to new record highs.
Recently, we have developed two approaches for the early forecasting of El Ni\~no. The climate network-based approach \cite{Ludescher2013, Ludescher2014, Bunde2024} allows forecasting the onset of an El Ni\~no event about 1 year ahead. The complexity-based approach \cite{Meng2019} allows additionally to forecast the magnitude of an upcoming El Ni\~no event in the calendar year before.
These methods successfully forecasted the onset of an Eastern Pacific El~Ni\~no for 2023/24 and the subsequent record-breaking warming of 2024 \cite{Ludescher2023a}.
Here, we apply these methods to forecast the ENSO state in 2025.
Both methods forecast the absence of an El~Ni\~no in 2025, with 91.2\% and 91.7\% probability, respectively.
Combining these forecasts with a logistic regression based on the Oceanic Niño Index (ONI) leads to a 69.6\% probability that 2025/26 will be a neutral ENSO event. We estimate the probability of a La~Ni\~na at 21.8\%. This makes it likely that the mean global temperature in 2025 will decrease somewhat compared to the 2024 level.
\end{abstract}

\section{The El~Ni\~no Southern Oscillation}

\begin{figure}[]
\begin{center}
\includegraphics[width=8cm]{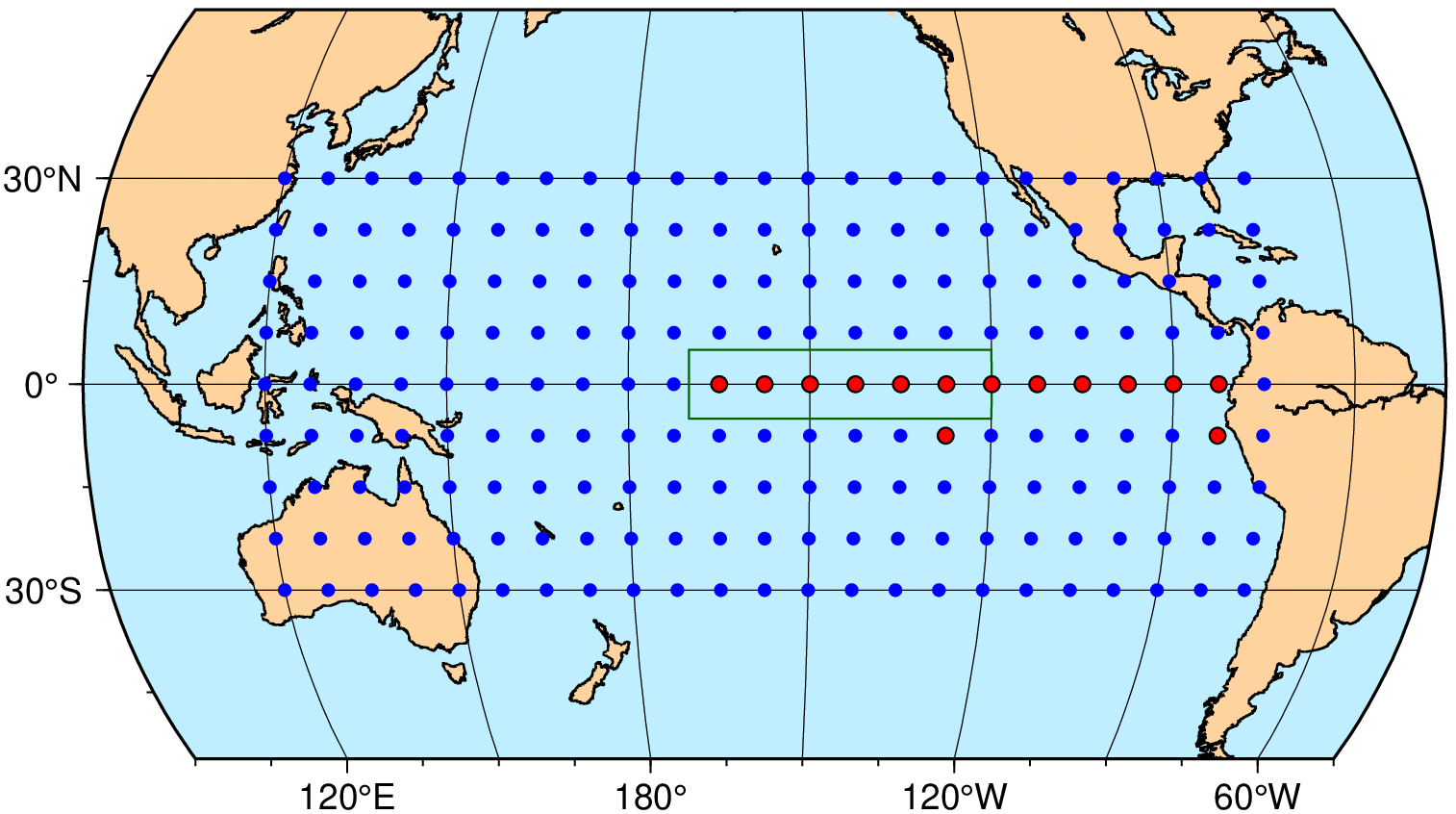}
\caption{{\bf The nodes of the climate network.} The network consists of 14 grid points in the central and eastern equatorial Pacific (red dots) and 193 grid points outside this area (blue dots). The green rectangle shows the Ni\~no3.4 area. The grid points represent the nodes of the climate network that we use here to forecast the {\it onset} or absence of an El~Ni\~no event. Each red node is linked to each blue node. The nodes are characterized by their surface air temperature (SAT), and the link strength between the nodes is determined from their cross-correlation (see below).}
\label{fig1}
\end{center}
\end{figure}

The El~Ni\~no-Southern Oscillation (ENSO) \cite{Clarke08, Sarachik10, Dijkstra2005, Wang2017,Timmermann2018, McPhadden2020} is a naturally occurring quasi-periodic oscillation of the Pacific ocean-atmosphere system that alternates between warm (El~Ni\~no), cold (La~Ni\~na) and neutral phases.
These phases are commonly defined via the sea surface temperature anomaly
(SSTA) in the Niño3.4 region (see Fig. \ref{fig1}). An El~Ni\~no is said to take place if the 3 months
running average SSTA in this region, i.e., the Oceanic Niño Index (ONI), is above or equal 0.5°C
for at least five consecutive months; correspondingly, below -0.5°C for a La~Ni\~na.
A regularly updated table of the ONI can be found at \cite{NOAA}.

Since El~Ni\~no and La~Ni\~na episodes can alter precipitation and temperature patterns over a large part of
the globe, which can result in, e.g., flooding or droughts \cite{Wen2002,Corral10,Donnelly07,Kovats03,Davis2001,McPhadden2020},
early-warning methods are highly desirable.
There are two general approaches to forecasting ENSO: (i) dynamical coupled
general circulation models (GCMs), which directly simulate the evolution of the system
according to the governing physical laws, e.g., the laws of fluid dynamics, and (ii) statistical
prediction models, which rely on statistical relationships between current properties/state of the
system and the system’s consequent development. These statistical relationships are derived
from past data, and the underlying physical processes are often not known.
Numerous models of both types have been proposed to forecast the pertinent index with lead times between 1 and 24 months
\cite{Cane86,Penland1995,Tziperman97,Kirtman03,Fedorov03,Galanti03,Chen04,Palmer2004,Luo08,Chen08,Chekroun11,Saha2014, Chapman2015, Lu2016, Feng2016,Rodriguez2016, Nootboom2018, Meng2018, Ham2019,DeCastro2020,Petersik2020,Hassanibesheli2022,Zhao2024,Zhao2024b,Schloer2024}.

Unfortunately, the current operational forecasts have quite limited anticipation power.
In particular, they generally fail to overcome the so-called ``spring barrier'' (see, e.g., \cite{Webster1995,Goddard2001}), which usually shortens their reliable warning time to around 6 months \cite{Barnston2012, McPhadden2020, Ehsan2024}.

To resolve this problem, we have recently introduced two alternative forecasting approaches, which considerably extend the probabilistic prediction horizon. The first approach \cite{Ludescher2013, Ludescher2014, Bunde2024} is based on complex network analysis \cite{Tsonis2006,Yamasaki2008,Donges2009,Gozolchiani2011,Dijkstra2019,Fan2020,Ludescher2021,Fan2022} and provides forecasts for the onset of an El~Ni\~no event, but not for its magnitude, in the year before the event starts.
The second approach \cite{Meng2019}
relies on the System Sample Entropy (SysSampEn), an information entropy, which measures the complexity (disorder) in the Ni\~no3.4 area. The method provides forecasts for both the onset and magnitude of an El Ni\~no event at the end of the previous year.
By regarding additional predictors, the El~Ni\~no forecasts of these methods can be leveraged to obtain more specific forecasts, for instance, for the type (Eastern Pacific or Central Pacific) of an El~Ni\~no event \cite{Ludescher2022,Ludescher2023b}.

The last El~Ni\~no forecast \cite{Ludescher2023a} based on data until December 2022 turned out to be correct. Both methods forecasted the onset of an El~Ni\~no, with a combined probability of around 89\%. The complexity-based approach forecasted a magnitude of $1.49\pm0.37$°C and the magnitude of the event turned out to be 2.0°C \cite{NOAA}, which is only 1.38 standard deviations from the center estimate.
Additionally, we had forecasted with an 87.5\% probability that an El~Ni\~no event starting in 2023 will be an Eastern Pacific event, as it turned out.

Here, we present the forecasts of both methods for 2025. Both methods forecast the absence of an El~Ni\~no in 2025, with 91.2\% and 91.7\% probability, respectively.
To discriminate between a La~Ni\~na and a neutral ENSO event, we use the interannual relationship of the ONI values as an additional predictor and obtain a $69.6\%$ probability for a neutral event in 2025.

\section{Climate network-based forecasting}

\subsection{The network-based forecasting algorithm}

For a brief description of the climate network-based approach we follow \cite{Ludescher2023a}.
The approach is based on the observation that a large-scale cooperative mode, linking the central and eastern equatorial Pacific with the rest of the tropical Pacific (see Fig. 1), builds up in the calendar year before an El~Ni\~no event.
According to \cite{Gozolchiani2011,Ludescher2013,Ludescher2014}, a measure for the emerging cooperativity can be derived from the time evolution of the teleconnections (``links``) between the surface air temperature anomalies (SATA) at the grid points (''nodes``) between the two areas. The strengths of these links are derived from the respective cross-correlations (for details, see, e.g., \cite{Ludescher2013,Ludescher2014}).

\begin{figure}[t!]
\begin{center}
\includegraphics[width=14cm]{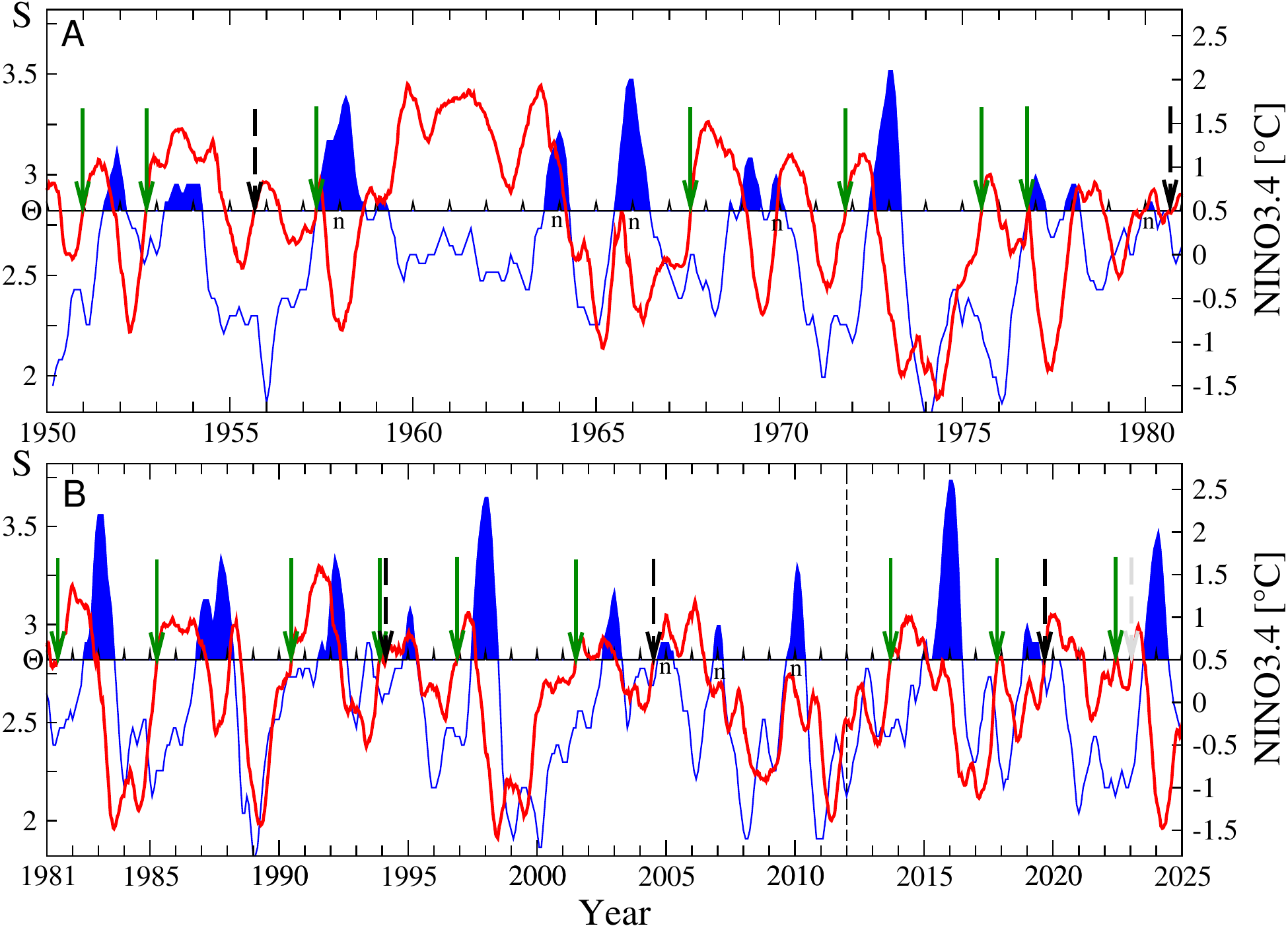}
\caption{{\bf The network-based forecasting scheme.}  We compare the average link strength $S(t)$
in the climate network (red curve) with a decision threshold $\Theta$ (horizontal line, here $\Theta = 2.82$), (left scale), and the standard Ni\~no3.4 index (ONI), (right scale), between January 1950 and December 2024.
When the link strength crosses the threshold from below, and the last available ONI is below $0.5^\circ$C,
we give an alarm and predict that an El~Ni\~no episode will start in the following calendar year.
The El~Ni\~no episodes (when the Ni\~no3.4 index is at or above $0.5^\circ$C for at least 5 months) are shown by the solid blue areas.
Correct predictions are marked by green arrows and false alarms by dashed arrows. The index $n$ marks not predicted events. The threshold learning phase is between 1950 and 1980. In the whole period between 1981 and December 2024, there were 12 El Ni\~no events. The algorithm generated 13 alarms, and 9 of them were correct. In the more restrictive version (ii) of the algorithm, only those alarms are considered where the ONI remains below 0.5°C for the rest of the year. In this version, the correct alarms in 1957 and 1976, and the incorrect alarms in 1994, 2004, 2019 and 2023 are not activated. Since the 2023 false alarm occurred after the introduction of version (ii) of the algorithm \cite{Ludescher2022}, it is shown as a dashed grey arrow. Between 1981 and December 2024, version (ii) of the algorithm gave 9 alarms, all of which were correct.
}
\label{fig2}
\end{center}
\end{figure}

The primary predictive quantity for the onset of an El~Ni\~no is the mean link strength $S(t)$ in the considered network, which is obtained by averaging over all individual links at a given time $t$ \cite{Ludescher2013,Ludescher2014}.
The mean link strength $S(t)$ typically rises in the calendar year before an El~Ni\~no event starts and drops with the onset of the event (see Fig. 2). This property serves as a precursor for the event. The forecasting algorithm involves as only fit parameter a decision threshold $\Theta$, which has been fixed in a learning phase (1950-1980) \cite{Ludescher2013}. Optimal forecasts in the learning phase are obtained for $\Theta$ between 2.815 and 2.834 \cite{Ludescher2013, Ludescher2014}.

The algorithm gives an alarm and predicts the onset of an El~Ni\~no event in the following year when $S(t)$ crosses $\Theta$ from below while the most recent ONI value is below $0.5^\circ$C. In a more restrictive version (ii) \cite{Ludescher2022, Ludescher2023b}, the algorithm considers only those alarms where the ONI remains below 0.5°C for the rest of the calendar year.

For the calculation of $S$, we use daily surface air temperature data from the National Centers for Environmental Prediction/National Center for Atmospheric Research (NCEP/NCAR) Reanalysis I project \cite{reanalyis1,reanalyis2}. We would like to note that for the calculations in the prediction phase (1981-present), e.g., of the climatological average, only data from the past up to the prediction date have been considered.

\subsection{El~Ni\~no forecasts since 2011}
The climate network-based algorithm has been quite successful in providing real-time forecasts, i.e., forecasts into the future. In its original version, it provided 12 forecasts for the period 2012-2023, 11 of these forecasts turned out to be correct, see Fig. 3.
The only incorrect forecast is a false alarm given in September 2019. In December 2022 \cite{Ludescher2022, Ludescher2023b}, we have introduced a more restrictive version (ii) of the algorithm, where only those alarms are considered, where the ONI remains below 0.5°C for the rest of the year and we are using exclusively this version from then on.
Version (ii) correctly forecasted the absence of an El~Ni\~no onset in 2024. Thus, in total, the method provided 13 real-time forecasts for the period 2012-2024, 12 of these forecasts turned out to be correct. The p-value, obtained from random guessing with the climatological El~Ni\~no onset probability \cite{Bunde2024}, for the skill in the forecasting period is $p\cong4.0\cdot10^{-3}$.
When considering the hindcasting and forecasting periods (1981-2024) correspondingly together, the p-value is $p\cong2.4\cdot10^{-5}$.

Throughout 2024, the mean link strength stayed well below the critical thresholds, thus predicting the absence of an El~Ni\~no onset in 2025. In the hindcasting and forecasting period (1981 and December 2024), there were 12 El~Ni\~no events. Version (i) of the algorithm generated 13 alarms and 9 of these were correct. Equivalently, the algorithm forecasted 30 times the absence of an El~Ni\~no onset and missed 3 El~Ni\~nos. Thus, forecasts for the absence of an El~Ni\~no are correct with 27/30 = 90\% probability.
The more restrictive version (ii) of the algorithm \cite{Ludescher2022, Ludescher2023b} gave 9 El~Ni\~no alarms, all of which were correct. In this version, the forecasts for the absence of an El~Ni\~no onset are correct with 31/34 $\approx$ 91.2\% probability.

\begin{figure}[]
\begin{center}
\includegraphics[width=11cm]{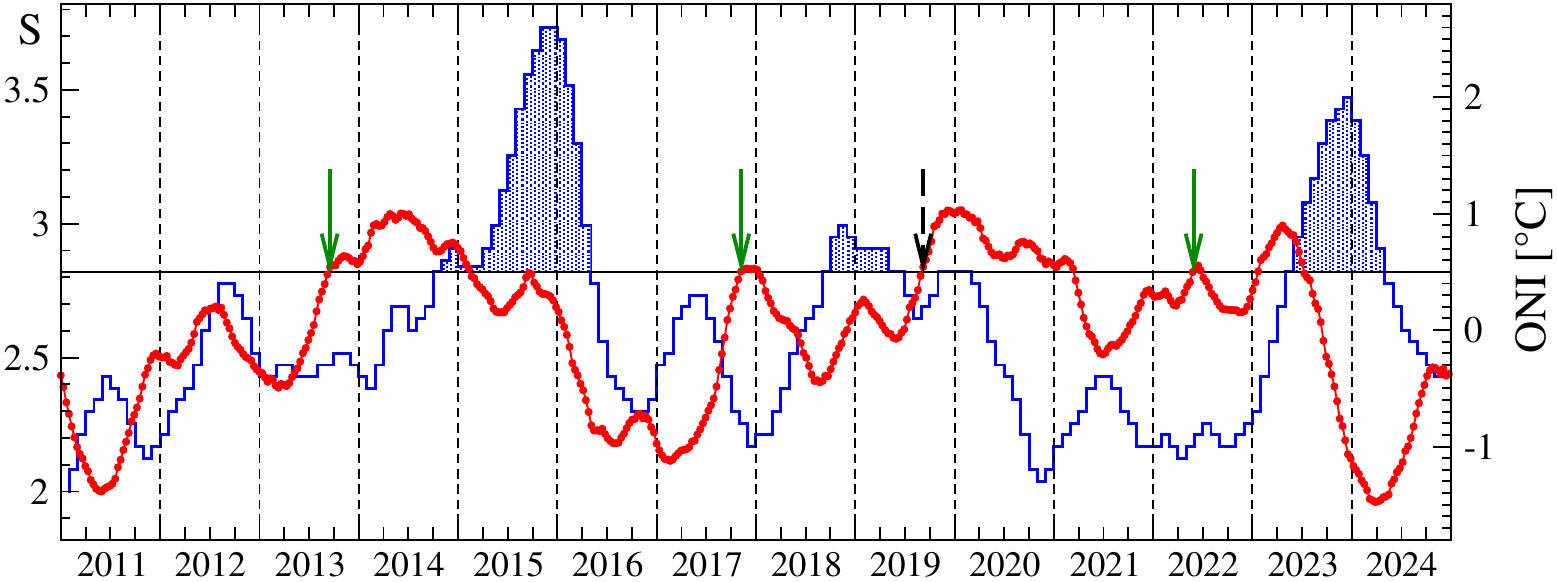}
\caption{ 
{\bf The climate network-based real-time forecasting phase.} Same as Fig. 2 but for the period between January 2011 and December 2024. The false alarm of version (i) of the algorithm in 2023 is not activated in version (ii) and is thus not shown here.
Throughout 2024, the average link strength $S(t)$ remained well below the critical threshold band, thus forecasting the absence of an El~Ni\~no in 2025.
}
\label{fig3}
\end{center}
\end{figure}

\section{System Sample Entropy-based forecast}

\subsection{SysSampEn} 

For a brief description of the System Sample Entropy (SysSampEn) approach we follow \cite{Ludescher2023a}.
The SysSampEn was introduced in \cite{Meng2019} as an analysis tool to quantify the complexity (disorder) in a complex system, in particular, in the temperature anomaly time series in the Ni\~no3.4 region.
For a brief description of the approach we follow \cite{Ludescher2023a}.
The SysSampEn is approximately equal to the negative natural logarithm of the conditional probability that 2 subsequences similar (within a certain tolerance range) for $m$ consecutive data points remain similar for the next $p$ points, where the subsequences can originate from either the same or different time series (e.g., black
curves in Fig. 4), that is,
\begin{equation}
SysSampEn(m, p, l_{eff}, \gamma) = -log(\frac{A}{B}),
\end{equation}
where A is the number of pairs of similar subsequences of length $m + p$, $B$ is the number of pairs of similar subsequences of length
$m$, $l_{eff} \leq l$ is the number of data points used in the calculation for each time series of length $l$, and $\gamma$ is a constant that determines the tolerance range. The detailed definition of the SysSampEn for a general complex system composed of $N$ time series and how to objectively choose the parameter values is described in detail in \cite{Meng2019}.

\begin{figure}[]
\begin{center}
\includegraphics[width=9.5cm]{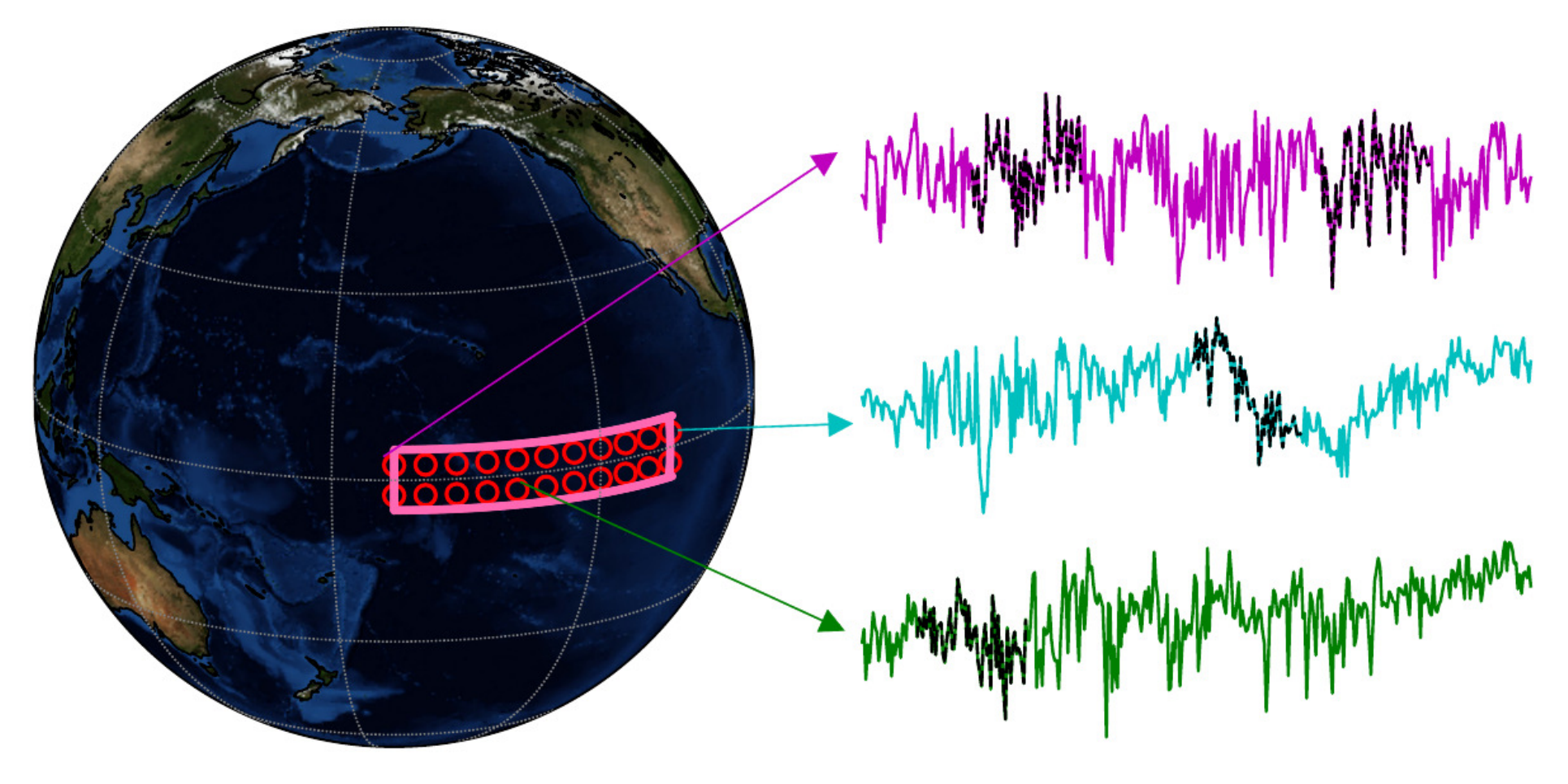}
\caption{{\bf The Ni\~no3.4 area and the SysSampEn input data.}
The red circles indicate the 22 nodes covering the Ni\~no 3.4 region at a spatial resolution of $5^\circ \times 5^\circ$.
The curves are examples of the temperature anomaly time series for 3 nodes in the Ni\~no 3.4 region for one specific year.
Several examples of their subsequences are marked in black. In the calculation of the SysSampEn, both the similarity of subsequences within a time series and the similarity of subsequences of different time series are regarded.
Figure from \cite{Meng2019}.}
\label{fig4}
\end{center}
\end{figure}

In \cite{Meng2019}, it was found the previous year's ($y-1$) SysSampEn exhibits a strong positive correlation ($r=0.90$ on average) with the magnitude of an El Ni\~no in year $y$ when parameter combinations are used that are able to quantify a system's complexity with a high accuracy. This linear relationship between SysSampEn and El Ni\~no magnitude enables thus to predict the magnitude of an upcoming El Ni\~no when the current ($y-1$) SysSampEn is inserted into the linear regression equation between the two quantities.

If the forecasted El~Ni\~no magnitude is below $0.5^\circ$C then the absence of an El Ni\~no onset is predicted for the following year $y$. Thus, SysSampEn values below a certain threshold indicate the absence of an El Ni\~no onset. In contrast, if the SysSampEn is above this threshold and the ONI in December of the current year is below $0.5^\circ$C then the method predicts the onset of an El~Ni\~no event in the following year.

The SysSampEn represents a generalization of two information entropies that are widely used tools in physiological fields: the sample entropy (SampEn) and the Cross-SampEn \cite{Richman2000}. For details about their relation to the SysSampEn, see \cite{Meng2019}.

\subsection{Forecast for 2025}

\begin{figure}[]
\begin{center}
\includegraphics[width=14.2cm]{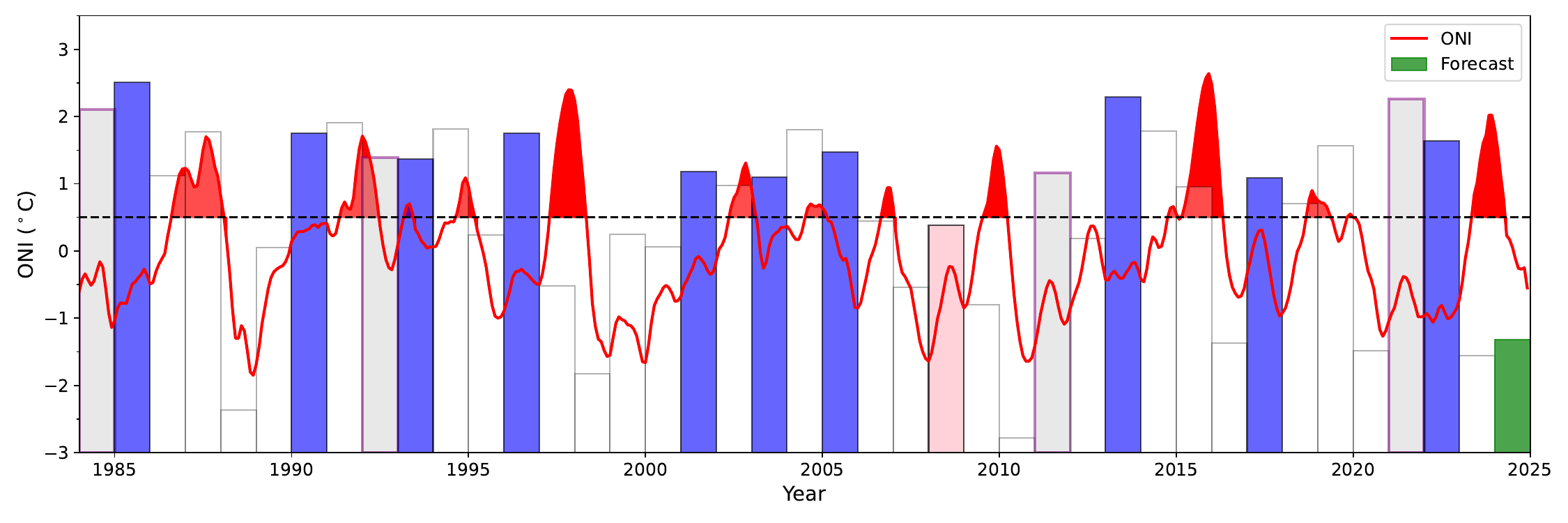}
\caption{{\bf Forecasted and observed El~Ni\~no magnitudes.} The magnitude forecast is shown as the height of rectangles in the year when the forecast is made, i.e., one year ahead of a potential El~Ni\~no. The forecast is obtained by inserting the regarded calendar year's SysSampEn value into the linear regression function between SysSampEn and El~Ni\~no magnitude. To forecast the following year's condition, we use the ERA5 daily near-surface (1000 hPa) temperatures with the set of SysSampEn parameters ($m = 30$, $p = 30$, $\gamma=8$ and $l_{eff} = 360$). The red curve shows the ONI and the red shades highlight the El~Ni\~no periods.
The blue rectangles show the correct prediction of an El~Ni\~no in the following calendar year.
The onset of an El~Ni\~no in the following year is predicted if the forecasted magnitude is above $0.5^\circ$C and the current year's December ONI is $<0.5^\circ$C. White dashed rectangles show correct forecasts for the absence of an El~Ni\~no.
Grey bars with a violet border show false alarms and the only missed event is shown as a pink rectangle.
The SysSampEn value for 2024 is $0.79$, i.e., is well below the threshold value of 1.31.
There were 12 occurrences of low SysSampEn accompanied by a low ONI
in December, as is the case in 2024 (green rectangle). In 11 out of these 12 cases, the hindcast was
correct. Thus the method predicts with 91.7\% probability the absence of an El~Ni\~no in 2025.
}
\label{fig5}
\end{center}
\end{figure}

Here we use as input data the daily near-surface (1000 hPa) air temperatures of the ERA5 reanalysis from the European Centre for Medium-Range Weather Forecasts (ECMWF) \cite{ERA5} analysed at a $5^\circ$  \ resolution. The last months in 2024 are from the initial data release ERA5T, which, in contrast to ERA5, only lags a few days behind real-time.

The daily time series are preprocessed by subtracting the corresponding climatological mean and then dividing by the climatological standard deviation. We start in 1984 and use the previous years to calculate the first anomalies.
For the calculation of the climatological mean and standard deviation, only past data up to the year of the prediction are used. For simplicity, leap days are excluded. We use the parameter values for the SysSampEn that lead to the best El~Ni\~no forecasting skill when applied to past events, as described in \cite{Meng2019}, $m = 30$, $p = 30$, $\gamma=8$ and $l_{eff} = 360$.

Figure 5 shows the results of the analysis. The magnitude forecast is shown as the height of rectangles in the year when the forecast is made, i.e., 1 year ahead of a potential El~Ni\~no onset. The forecast is obtained by inserting the regarded calendar year's SysSampEn value into the linear regression function between SysSampEn and El~Ni\~no magnitude. For the 2025 forecast, the regression is based on all correctly hindcasted El~Ni\~no events before 2024. The red curve shows the ONI and the red shades indicate the El~Ni\~no periods. The blue rectangles show the correct prediction of an El~Ni\~no in the following calendar year and grey rectangles with a violet border show false alarms.
White dashed rectangles show correct forecasts for the absence of an El~Ni\~no and the only missed event is the 2009/10 El~Ni\~no, where the preceding SysSampEn value was slightly below the threshold.

There were 12 occurrences of a low SysSampEn accompanied by a lower
than 0.5°C ONI in December. In 11 out of these 12 cases, the hindcast was correct. The forecasted
magnitude for 2025 is far below 0.5°C, as shown by the green rectangle. The SysSampEn value
for 2024 is 0.79, i.e., well below the threshold value of 1.31. Thus the method predicts with 91.7\%
probability the absence of an El~Ni\~no in 2025.

\section{Probability of a La~Ni\~na vs. a neutral event in 2025}

\begin{figure}[]
\begin{center}
\includegraphics[width=10cm]{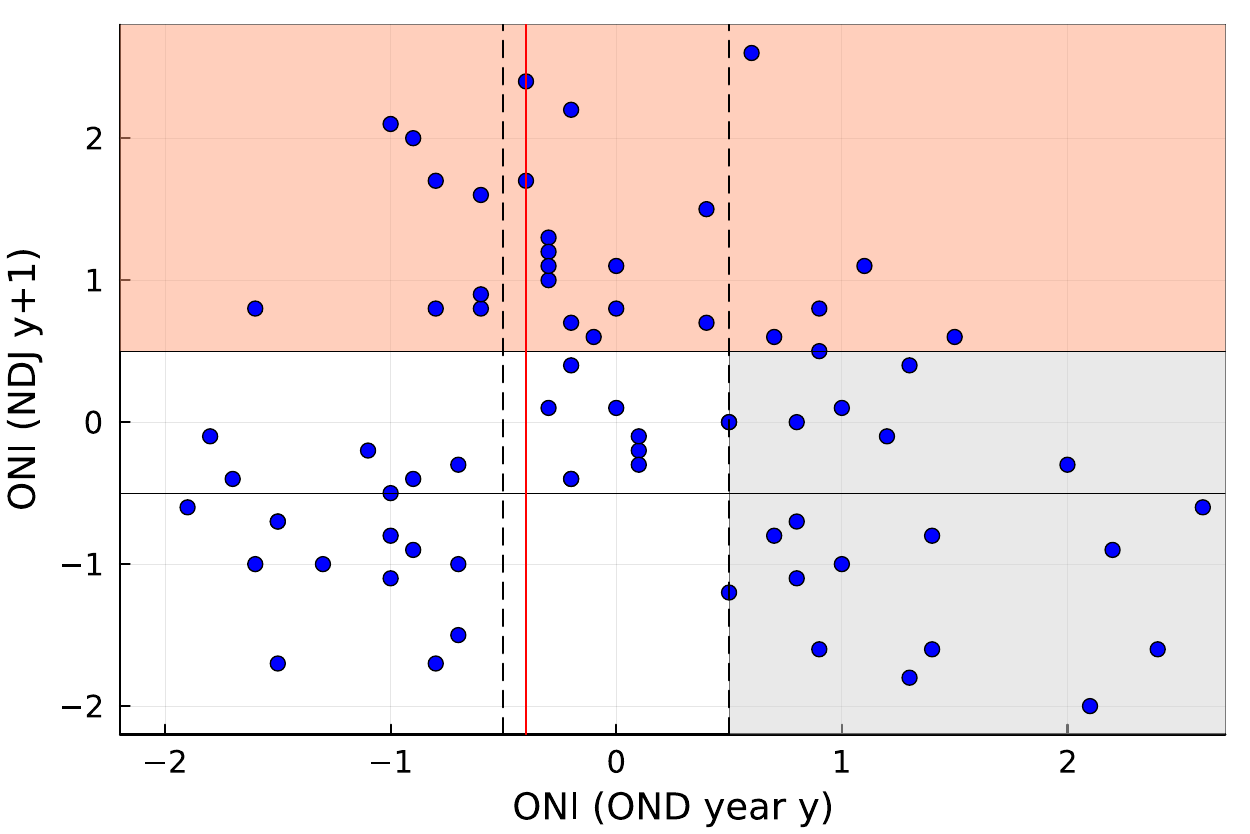}
\caption{{\bf Interannual ONI relationship.} The November-December-January (NDJ) ONI value of year y+1 vs. the OND value of the previous calendar year y (blue circles). Our forecast predicts the absence of an El~Ni\~no with a high probability (91.2\% and 91.7\%). Thus, this outcome is excluded with a high probability and is shown as a light red area. Dashed vertical black lines mark -0.5°C and 0.5°C and the current OND value -0.4°C is shown as a vertical red line. Since we are currently clearly not in an El~Ni\~no event, this excludes additionally the grey area.}
\label{ONI_interannual}
\end{center}
\end{figure}

Both the climate network-based and the complexity-based methods forecast with a high probability the absence of an El~Ni\~no event in 2025. The  respective probabilities are 91.2\% and 91.7\%, and the combined probability is 91.4\%.  To further specify our forecast and to discriminate between a La~Ni\~na or a neutral event, we analyze the relationship between interannual ONI values. Figure \ref{ONI_interannual} shows the NDJ ONI values of year y+1 vs the OND value of year y.
We use NDJ since El~Ni\~no and La~Ni\~na tend to peak around December, and OND to gain an additional month of lead time.
The forecasted absence of an El~Ni\~no in 2025, by our methods, excludes (with high probability) a large area (red shading).
The current OND ONI value in 2024 is -0.4°C, i.e., a neutral event, and is shown as a vertical red line. Since we are currently clearly not in an El~Ni\~no event, this excludes additionally the grey area. We regard an OND ONI value of 0.5°C as part of an El~Ni\~no since, in 2 out of 3 past cases, this was the eventual turnout, and whether such a weak warming is part of an El~Ni\~no is not known in real-time. The exception was the 2019/2020 warming event, which narrowly missed satisfying the definition of an El~Ni\~no.

Figure \ref{ONI_interannual} shows that in the observational record since 1950, neutral events were not followed by a La~Ni\~na phase. This allows a first rough estimate of the neutral vs. La~Ni\~na probability. There were 8 cases (in Figure \ref{ONI_interannual}, 2 such cases are on top of each other) where a neutral event was followed by a second neutral event. Assuming that no El~Ni\~no will start in 2025, a neutral event remains thus the only outcome that has been observed since 1950. Applying a Bayesian-type estimate using Laplace's rule of succession \cite{Jaynes2003}, we arrive at a neutral event probability of $(8+1)/(8+2)=0.90$. Correspondingly, the probability of a La~Ni\~na is $0.1$.

\begin{figure}[]
\begin{center}
\includegraphics[width=10cm]{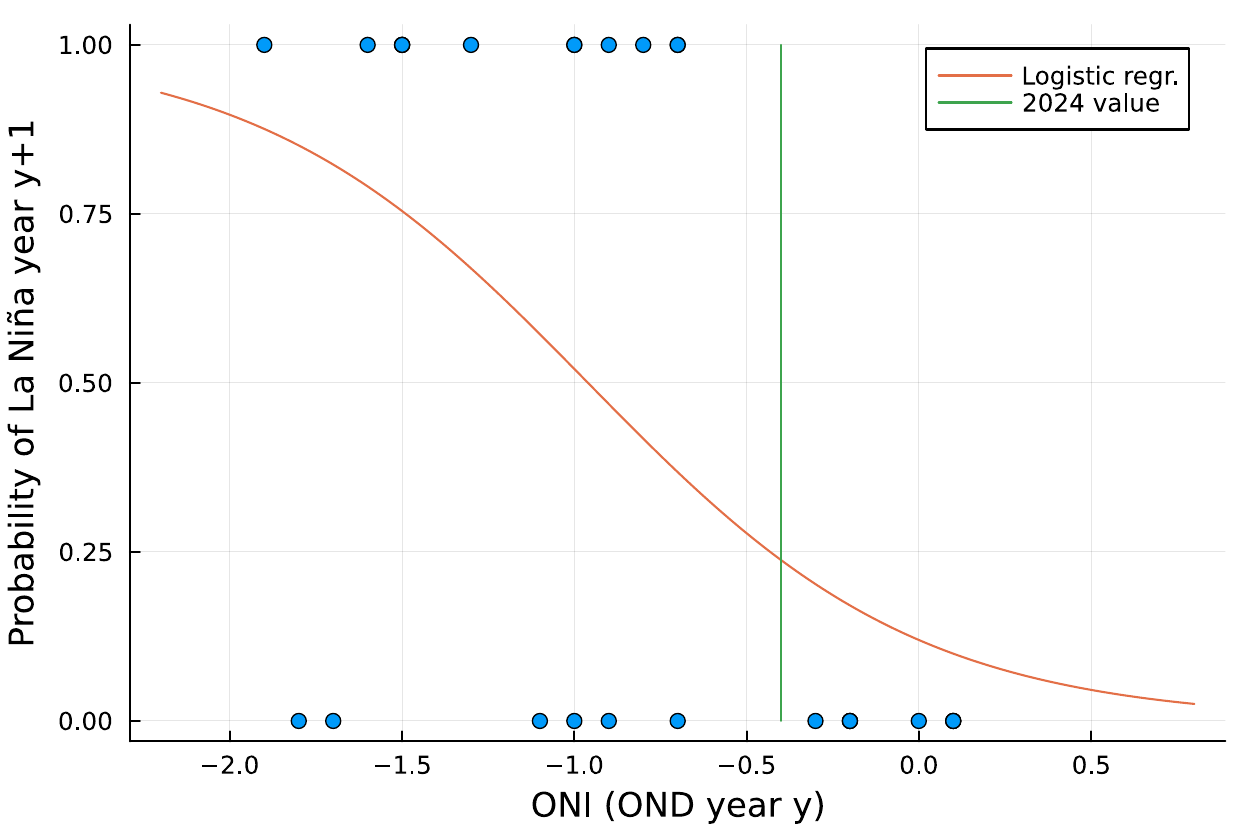}
\caption{{\bf The probability of a La~Ni\~na}. We regard all years that are non-El~Ni\~no years and are also followed by a non-El~Ni\~no year, i.e., the events shown in the white area in Figure \ref{ONI_interannual}. These events correspond to the current state since the OND ONI is $-0.4$°C and our forecasting methods predict the absence of an El~Ni\~no in 2025 with high probability. The outcome of the second year is encoded as 1 for La~Ni\~na and 0 for a neural event (blue circles). To obtain the probability for a La~Ni\~na event in 2025, given that no El~Ni\~no starts in 2025, we apply a logistic regression and use the current OND ONI value of $-0.4$°C. Excluding El~Ni\~no, we obtain a $23.8\%$ probability for a La~Ni\~na vs. a $76.2\%$ probability for a neutral event.
}
\label{fig:log-reg}
\end{center}
\end{figure}

To take full advantage of the available data that is relevant in the current case, we regard all years that are non-El~Ni\~no (i.e., La~Ni\~na or neutral) years and are also followed by a non-El~Ni\~no year. These events are shown in the white area in Figure \ref{ONI_interannual}.
On this data, we perform a logistic regression with the outcomes La~Ni\~na (encoded as 1) and neutral (encoded as 0), see Figure \ref{fig:log-reg}. We use as a predictor the OND ONI value of the year $y$, which is preceding the target year $y+1$. Based on this regression and the current OND ONI value of -0.4°C, we obtain a $23.8\%$ probability for a La~Ni\~na and a $76.2\%$ probability for a neutral ENSO year.
To take into account that there is a 0.086, i.e., $8.6\%$, probability for an El~Ni\~no in 2025 when combining the climate network and the complexity-based approaches, we multiply the above values with the complementary probability of $0.914$ and arrive at $21.8\%$ and $69.6\%$ overall probabilities for a La~Ni\~na and a neutral event, respectively.

\begin{figure}[]
\begin{center}
\includegraphics[width=10cm]{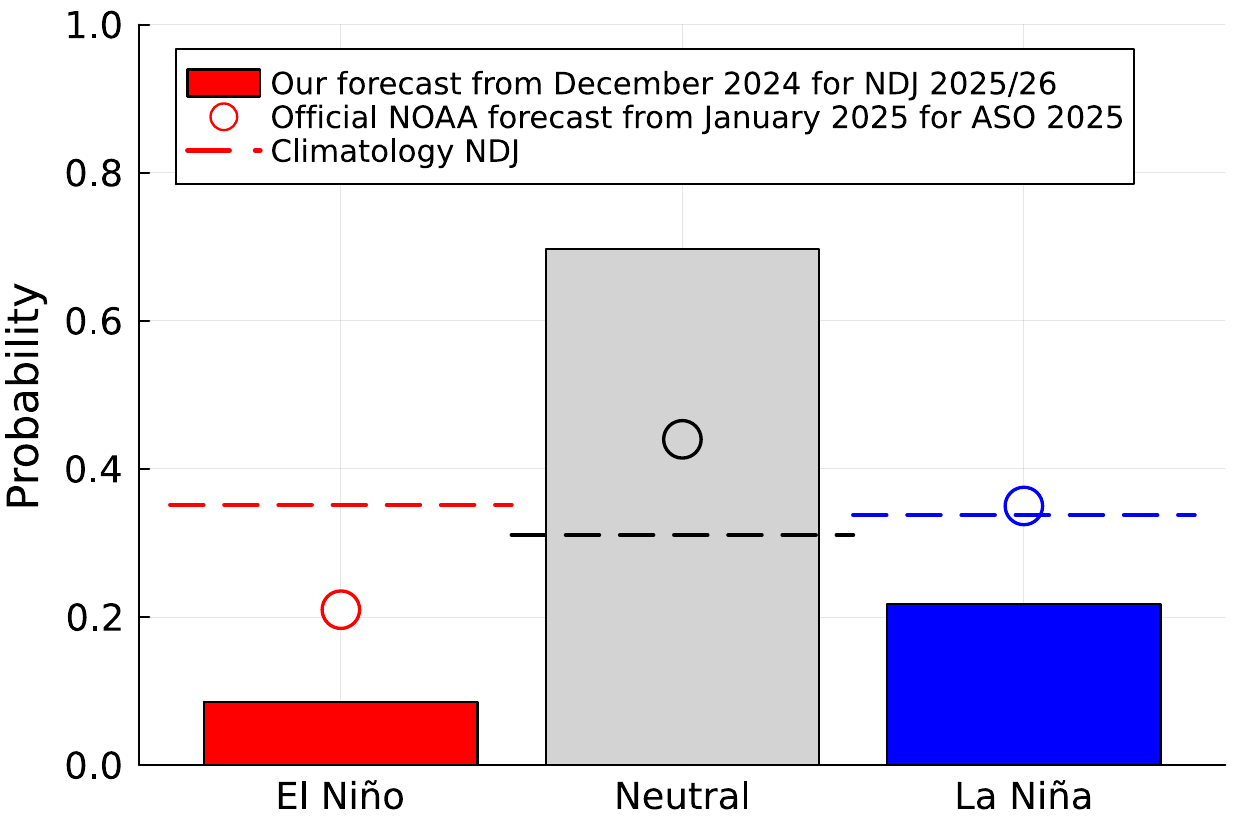}
\caption{{\bf Summary of our forecast.} The height of the bars shows the probability of El~Ni\~no, neutral event and La~Ni\~na for NDJ 2025/26 based on the climate network, the SysSampEn and on a logistic regression based on the events that correspond to the current state. We obtain $8.6\%$, $69.6\%$ and $21.8\%$, respectively. The corresponding probabilities of the NOAA CPC forecast from January 2025 are $21\%$, $44\%$ and $35\%$ for the target season ASO 2025 \cite{CPC}. The dashed horizontal lines show the NDJ climatological probabilities.
}
\label{fig:forecast}
\end{center}
\end{figure}

Figure \ref{fig:forecast} summarizes our final forecast probabilities for NDJ 2025/26 and compares them with the climatologically expected outcome (dashed horizontal lines) and the current official NOAA climate prediction center (CPC) probabilistic forecast for August-September-October (circles) issued in January 2025 \cite{CPC}.
Since El~Ni\~nos and La~Ni\~nas can develop later in the calendar year, the NOAA CPC forecast implies that for NDJ 2025/26, there is a lower probability of a neutral event than the provided probability for ASO. Thus, compared to the NOAA CPC forecast, we predict with a considerably higher probability a neutral ENSO year.

\newpage

\section*{Acknowledgements}
J. L. is part of the Planetary Boundaries Science Lab's research effort at PIK.

\end{document}